\documentclass[aps,prb,twocolumn,superscriptaddress,showpacs]{revtex4}
\usepackage{graphicx}
\usepackage{amssymb}
\usepackage{amsfonts}
\usepackage{bm}
\begin{document}

\title{Origin and control of ferromagnetism in dilute magnetic semiconductors and oxides}

\author{Tomasz Dietl}
\affiliation{Institute of Physics, Polish Academy of Science,
al.~Lotnik\'ow 32/46, PL 02-668 Warszawa, Poland}
\affiliation{Institute of Theoretical Physics, Warsaw University,
PL 00-681 Warszawa, Poland} \affiliation{ERATO Semiconductor
Spintronics Project, Japan Science and Technology Agency,
al.~Lotnik\'ow 32/46, PL 02-668 Warszawa, Poland}

\date{\today}

\begin{abstract}
The author reviews the present understanding of the hole-mediated ferromagnetism in magnetically doped semiconductors and oxides as well as the origin of high temperature ferromagnetism in materials containing no valence band holes.  It is argued that in these systems spinodal decomposition into regions with a large and a small concentration of magnetic component takes place. This self-organized assembling of magnetic nanocrystals can be controlled by co-doping and growth conditions. Functionalities of these multicomponent systems are described together with prospects for their applications in spintronics, nanoelectronics, photonics,  plasmonics, and thermoelectrics.

\end{abstract}

\pacs{75.50.Pp}

\maketitle

\section{Introduction}

The demonstrated and foreseen functionalities of ferromagnetic semiconductors such as (Ga,Mn)As,\cite{Dietl:2007_b}  together with theoretical promises,\cite{Dietl:2000_a} have triggered a considerable endeavor aiming at developing semiconductor systems in which spontaneous magnetization would persist to above room temperature. Comprehensive experimental efforts have been stimulated further by results of ample first-principles computations which have substantiated the possibility of a robust ferromagnetism in a variety of systems,\cite{Sato:2002_a,Sandratskii:2003_a} even in the absence of valence-band holes, which were originally regarded as a necessary ingredient for high temperature ferromagnetism in magnetically doped semiconductors.\cite{Dietl:2000_a} Indeed, a ferromagnetic response persisting up to  above room temperature has been reported for a broad class of diluted magnetic semiconductors (DMS) and diluted magnetic oxides (DMO) containing minute amounts of magnetic ions.\cite{Pearton:2003_a,Fukumura:2005_a,Liu:2005_a,MacDonald:2005_a,Chambers:2006_a}

However, this fast and promising development of high-temperature ferromagnetic semiconductors has been challenged on both experimental and theoretical fronts. In particular, it has become more and more clear that the adequate experimental depiction of these systems requires the application of element-specific characterization tools with nanoscale spatial resolution. On the theoretical side, it has been appreciated that conceptual difficulties of charge transfer insulators and strongly correlated disordered metals are combined in these materials with intricate aspects of heavily doped semiconductors and semiconductor alloys, such as Anderson-Mott localization, defect formation by self-compensation mechanisms, spinodal decomposition, and the breakdown of the virtual crystal approximation. This blend of intricacies is beyond capabilities of the standard {\em ab initio} approaches involving supercells and the local spin density approximation (LSDA) or its variants.

In this paper, the recent progress in  understanding of the origin of ferromagnetism in DMS and DMO is reviewed. It is argued that in the case of a random distribution of magnetic ions, delocalized or weakly localized holes in the valence band are necessary to promote ferromagnetism characterized by a reasonably high Curie temperature, say $T_{\mathrm{C}} \gtrsim 20$~K. Interestingly, however, when the concentration of magnetic impurities exceeds the solubility limit, a variety of magnetic nanocrystals embedded coherently into the semiconductor are formed in a self-organized manner. According to the present insight, these nanocrystals containing a high density of the magnetic constituent account for the persistence of ferromagnetic features up to high-temperatures. Importantly, the shape and dimension, and thus the blocking temperature of the nanocrystals can be controlled by co-doping with shallow dopants and through growth conditions. Appealing functionalities of these multi-component systems are presented and prospects for their applications are listed in the last section of the paper.

\section{Hole mediated ferromagnetism - weak coupling limit}

There is a general consensus now that ferromagnetism of Mn-doped III-V antimonides, arsenides, and phosphides as well as of II-VI tellurides is associated with the presence of holes. However, the question much discussed recently is whether the holes mediating the ferromagnetic coupling, particularly in the model system (Ga,Mn)As, reside in the valence band or in an impurity band detached from the valence band.\cite{Jungwirth:2007_a,Ohno:2007_a} In this section, we recall experiments described within the $p$-$d$ Zener model, in which the holes are assumed to occupy the valence band. We also quote some difficulties in the description of experimental data within the impurity band picture.

\subsection{Disorder-free $p$-$d$ Zener model}

The $p$-$d$ Zener model of hole-mediated ferromagnetism has been developed in the spirit of the second-principles method (as opposed the first-principles computations) which exploits, whenever possible, experimental information on  pertinent material characteristics, such as the host band structure, the ground state of the relevant magnetic impurities, and the coupling between these two subsystems. In particular, the host band structure can be accurately parameterized  by the multi-band $kp$ method,\cite{Dietl:2001_b,Hankiewicz:2004_c} the multi-orbital tight-binding approximation,\cite{Sankowski:2005_a,Timm:2005_b,Sankowski:2007_a} or their combination.\cite{Vurgaftman:2001_a}

Within the disorder-free $p$-$d$ Zener model, the translational symmetry is restored by employing the virtual-crystal and molecular-filed approximations, which constitute the experimentally verified approach for the description of semiconductor alloys such as (Al,Ga)As and magnetic alloys, for example (Cd,Mn)Te.\cite{Dietl:1994_a} The $p$-$d$ Zener model has been successful in explaining a number of properties observed in ferromagnetic DMS, in particular (Ga,Mn)As and (In,Mn)As. The properties are the ferromagnetic transition temperature $T_{\mathrm{C}}$,\cite{Dietl:2000_a,Dietl:2001_b,Jungwirth:2006_c} magnetization and effective Land\'e factor,\cite{Sliwa:2006_a} Gilbert constant,\cite{Sinova:2004_b} spin-polarization of the hole liquid,\cite{Dietl:2001_b,Barden:2003_a} magnetocrystalline anisotropy with its strain and temperature dependence,\cite{Dietl:2000_a,Dietl:2001_b,Sawicki:2004_d,Takamura:2002_a} magnetic stiffness,\cite{Konig:2001_a,Potashnik:2002_a} stripe domain width,\cite{Dietl:2001_c} the anomalous Hall effect,\cite{Jungwirth:2006_a} magnetic anisotropy of the Coulomb blockade,\cite{Wunderlich:2006_a} and optical properties including the peculiar temperature dependence of the band-edge behavior of the magnetic circular dichroism.\cite{Dietl:2001_b}

While the $kp$ method is the approach of choice for films, the tight-binding scheme is better suited for treating multi-layer structures. Such a theory,\cite{Sankowski:2005_a,Sankowski:2007_a} in which $sp^3d^5s^*$ orbitals are considered, takes care about effects of interfaces and inversion symmetry breaking as well as the band dispersion in the entire Brillouin zone, so that the essential for the spin-dependent tunneling Rashba and Dresselhaus terms as well as the tunneling \emph{via} $\vec{k}$ points away from the zone center are properly taken into account. The approach in question, provides information on the sign and magnitude of the interlayer coupling\cite{Sankowski:2005_a} and, supplemented by the Landauer-B\"uttiker formalism, explains experimental magnitudes of electron spin current polarization  in Esaki-Zener diodes\cite{Dorpe:2005_a}  and tunneling magnetoresistance (TMR) in trilayer structures.\cite{Sankowski:2007_a} Furthermore, theory reproduces a fast decrease of spin-polarization and TMR with the device bias as well their dependence on the magnetization direction.\cite{Sankowski:2007_a,Dorpe:2005_a}

\subsection{Effects of disorder}

Many experiments show that hole-controlled DMS ferromagnetic films reside on or at the vicinity of the metal-insulator transition boundary,\cite{Matsukura:1998_a,Jungwirth:2007_a,Sheu:2007_a,Scarpulla:2005_a} indicating that both disorder and carrier-carrier correlations are important in these systems.  A rather strong disorder originates from Coulomb potentials of acceptors, that supply holes, and donors, whose concentration is usually enhanced by a self-compensation mechanism, in addition to the local potentials introduced by magnetic impurity cores, containing spin-dependent (exchange) and spin-independent (chemical shift) contributions. As already mentioned, the question of whether the holes accounting for ferromagnetism of (Ga,Mn)As occupy an impurity band or reside in the GaAs-like valence band is a subject of the recent dispute.\cite{Jungwirth:2007_a,Ohno:2007_a}

The Curie temperature $T_{\mathrm{C}}$ of (Ga,Mn)As shows no critical behavior at the metal-insulator transition\cite{Matsukura:1998_a} but it vanishes rather rapidly when moving away from the metal-insulator transition into the insulator phase.\cite{Matsukura:1998_a,Sheu:2007_a} At the same time, $T_{\mathrm{C}}$ grows steadily with the magnitude of the conductivity on the metal side of the metal-insulator transition.\cite{Matsukura:1998_a,Potashnik:2001_a,Yu:2002_a,Campion:2003_b} Guided by these observations, the band scenario has been proposed in order to describe the ferromagnetism in magnetic III-V semiconductors on both sides of the metal-insulator transition.\cite{Dietl:2000_a,Dietl:2001_b}  Within this model, the hole localization length, which diverges at the metal-insulator transition, remains much greater than the average distance between acceptors for the experimentally important range of hole densities.  Thus, the holes can be regarded as delocalized at the length scale relevant for the coupling between magnetic ions. The spin-spin exchange interactions are effectively mediated by the itinerant carriers, so that the $p$-$d$ Zener model can be applied also to the insulator side of the metal-insulator transition.

\subsection{Impurity-band models}

Because of the importance of localization phenomena, variety of ferromagnetism models have been proposed -- but not yet quantitatively compared with experimental findings -- which assume that in the relevant hole concentration range, the impurity and valence bands have not yet merged, so that the Fermi level actually resides within the impurity band in (Ga,Mn)As.\cite{Sheu:2007_a,Inoue:2000_a,Litvinov:2001_a,Berciu:2001_a,Kaminski:2002_a} We note that the generic expectation of all the impurity models is a maximum of $T_{\mathrm{C}}$ at a half filling of the impurity band.\cite{Popescu:2007_a} This prediction is at variance with the available experimental data which show a monotonous increase of the $T_{\mathrm{C}}$ value as compensation diminishes to zero.\cite{Jungwirth:2006_c,Potashnik:2001_a,Scarpulla:2005_a} In contrast, when the carrier density becomes greater that the magnetic impurity concentration, a decrease of the $T_{\mathrm{C}}$ magnitude is expected as a result of the RKKY-like oscillations.\cite{Dietl:2001_b,Timm:2005_b}

\section{Strong coupling limit}

An interesting question arises about the accuracy of virtual-crystal (VCA) and molecular-field approximations (MFA). A moderate magnitude of the Mn acceptor binding energy together with the experimentally estimated the GaAs/(Ga,Mn)As band offset and the $p$-$d$ exchange energy indicate that VCA and MFA are valid in (Ga,Mn)As, similarly to the case of the corresponding II-VI alloys, such as (Cd,Mn)Te and (Zn,Mn)Se. However, it has recently been shown that this may not be the case in magnetically doped III-V nitrides and II-VI oxides, where a large $p$-$d$ hybridization makes the core potential of transition metal impurities to be strong enough to bind a hole.\cite{Dietl:2007_c} This makes these two families of DMS similar to other non-VCA alloys, such as Ga(As,N).\cite{Wu:2002_a} A number of surprising effects is observed in such a case, including the sign reversal of the valence-band splitting accompanied by a reduction in its magnitude.\cite{Pacuski:2007_a}

The strong coupling shifts the metal-insulator transition to higher hole concentrations than those attained in samples studied till now. According to our insight,\cite{Dietl:2000_a,Dietl:2001_b} in the absence of delocalized or weakly localized holes, no ferromagnetism is expected for randomly distributed diluted spins. Indeed, recent studies of (Ga,Mn)N indicate that in samples containing up to 6\% of Mn, holes stay strongly localized and, accordingly, $T_{\mathrm{C}}$ below 10~K is experimentally revealed.\cite{Edmonds:2005_a,Sarigiannidou:2006_a} An interesting intermediate situation is realized in (Ga,Mn)P, where $T_{\mathrm{C}}$ reaches 60~K.\cite{Scarpulla:2005_a}
However, at sufficiently high hole densities, an insulator-to-metal transition is expected. In the metallic phase, many-body screening of local potentials may annihilates bound states. Large spin-splitting and robust ferromagnetism are
expected in this regime.\cite{Dietl:2002_a,Popescu:2007_a}

\section{Non-uniform ferromagnetic DMS}

So far we have assumed a random but macroscopically uniform distribution of magnetic impurities. Actually, the question about the relation between high $T_{\mathrm{C}}$ and the magnetic ion distribution has been around for many years.\cite{Dietl:2003_a} In particular, as pointed out elsewhere,\cite{Dietl:2005_c} both hexagonal and zinc-blende Mn-rich (Ga,Mn)As nanocrystals are clearly seen in annealed (Ga,Mn)As.\cite{Moreno:2002_a,Yokoyama:2005_a} These two cases correspond to precipitation of other phases and spinodal decomposition, respectively.  As emphasized in recent reviews,\cite{Dietl:2005_c,Katayama-Yoshida:2007_a,Dietl:2007_a,Bonanni:2007_b} the film decomposition into nanoregions with low and high concentrations of magnetic ions constitutes a generic property of many DMS and DMO, occurring if the mean density of the magnetic constituent exceeds the solubility limit at given growth or thermal processing conditions. Importantly, the host can stabilize magnetic nanostructures in a new crystallographic and/or chemical form,  not as yet listed in materials compendia.

\subsection{Observations of spinodal decomposition}

It is well known that phase diagrams of a number of alloys exhibit a solubility gap in a certain concentration range. This may lead to a spinodal decomposition referred to above. If the concentration of one of the alloy constituents is small, it may appear in a form of coherent nanocrystals embedded by the majority component. For instance, such a spinodal decomposition occurs in (Ga,In)N,\cite{Farhat:2002_a} where In-rich quantum-dot like regions are embedded by the In poor matrix. However, according to the pioneering {\em ab initio} work of van Schilfgaarde and Mryasov\cite{Schilfgaarde:2001_a} and others\cite{Sato:2005_a,Kuroda:2007_a} particularly strong tendency to form non-random alloy occurs in the case of many DMS. For instance, the evaluated gain in energy by bringing two Ga-substitutional Mn atoms together is $E_{\mathrm{d}} = 120$~meV in GaAs and 300~meV in GaN, and reaches 350~meV in the case of Cr pair in GaN.\cite{Schilfgaarde:2001_a}

Since spinodal decomposition does not usually involve any precipitation of other crystallographic phases, it is not easy detectable experimentally. Nevertheless, its presence was found by electron transmission microscopy (TEM) in (Ga,Mn)As,\cite{Moreno:2002_a,Yokoyama:2005_a}  where coherent zinc-blende Mn-rich (Mn,Ga)As metallic nanocrystals led to the apparent Curie temperature up to 360~K.\cite{Yokoyama:2005_a} Furthermore, coherent hexagonal nanocrystals were detected by spatially resolved x-ray diffraction in (Ga,Mn)N\cite{Martinez-Criado:2005_a} and by TEM with energy dispersive spectroscopy (EDS) in (Al,Cr)N, (Ga,Cr)N,\cite{Gu:2005_a} and (Ga,Fe)N.\cite{Bonanni:2007_a} The same technique revealed spinodal decomposition in (Zn,Cr)Te,\cite{Kuroda:2007_a,Sreenivasan:2007_a} confirming indirectly the early suggestion that the presence of ferromagnetism in (Zn,Cr)Se points to a non-random Cr distribution.\cite{Karczewski:2003_a} Finally, we mention the case of (Ge,Mn). For this DMS, one group\cite{Bougeard:2006_a} observed Mn-rich regions in the form of nanodots of diameter of 5~nnm, whereas another team\cite{Jamet:2006_a} found periodically arranged nanocolumns of diameter of 2~nm, which extended from the substrate to the surface. Actually, a tendency for the nanocolumn formation was also reported for (Al,Cr)N.\cite{Gu:2005_a}  This demonstrates that growth conditions can serve to control nanocrystal shapes. Interestingly, these two kinds of nanocrystal forms were reproduced by Monte-Carlo simulations.\cite{Katayama-Yoshida:2007_a}

Since magnetic nanocrystals assembled by spinodal decomposition assume the form imposed by the matrix, it is {\em a priori} unknown of whether they will be metallic or insulating as well as of whether they will exhibit ferromagnetic, ferrimagnetic or antiferromagnetic spin order. However, owing to the large concentration of the magnetic constituent within the nanocrystals, their spin ordering temperature is relatively high, typically above the room temperature. It is obvious that ferromagnetic or ferrimagnetic nanocrystals will lead to spontaneous magnetization and magnetic hysteresis up to the blocking temperature.\cite{Shinde:2004_a,Goswami:2005_a} Interestingly,  uncompensated spins at the surface of antiferromagnetic nanocrystals can also result in a sizable value of spontaneous magnetization up to the usually high N\'eel temperature.\cite{Eftaxias:2005_a} It has recently been suggested\cite{Dietl:2007_d} that the ferromagnetic-like behavior of (Zn,Co)O,\cite{Chambers:2006_a,Dietl:2007_d} in which no precipitates of other crystallographic phases are detected, originates from the presence of wurtzite antiferromagnetic clusters of CoO or Co-rich (Zn,Co)O.  Such nanocrystals can aggregate under appropriate growth conditions or thermal treatment.

It is, therefore, legitimate to suppose that coherent nanocrystals with a large concentration of the magnetic constituent and a sufficiently large magnitude of magnetic anisotropy account for high apparent Curie temperatures detected in a number of DMS and DMO. This model explains a long staying puzzle about the origin of ferromagnetic-like response in DMS and DMO, in which the average concentration of magnetic ions is below the percolation limit for the nearest neighbor coupling and, at the same time, the itinerant hole density is too low to mediate an efficient long-range exchange interaction. Remarkably, the presence of magnetically active metallic nanocrystals leads to enhanced magnetotransport\cite{Jamet:2006_a,Shinde:2004_a,Ye:2003_a,Kuroda:2007_a} and magnetooptical\cite{Yokoyama:2005_a,Kuroda:2007_a} properties over a wide spectral range. This opens doors for various applications of such hybrid systems provided that methods for controlling nanocrystal characteristics and distribution would be elaborated.

\subsection{Controlling spinodal decomposition by inter-ion Coulomb interactions}

So far, the most rewarding method of controlling the self-organized growth of coherent nanocrystals or quantum dots has exploited strain fields generated by lattice mismatch at interfaces of heterostructures.\cite{Stangl:2004_a}  Remarkably, it becomes possible to fabricate highly ordered three dimensional dot crystals under suitable spatial strain anisotropy.\cite{Stangl:2004_a}

It has been suggested that outstanding properties of magnetic ions in semiconductors can be utilized for the control of self-organized assembly of magnetic nanocrystals:\cite{Dietl:2006_a} It is well know that in most DMS and DMO the levels derived from the open $d$ or $f$ shells of magnetic ions reside in the band gap of the host semiconductor. This property of magnetic ions has actually been exploited for a long time to fabricate semi-insulating materials, in which carriers introduced by residual impurities or defects are trapped by the band-gap levels of magnetic impurities. The relevant observation here is that such a trapping alters the charge state of the magnetic ions and, hence, affect their mutual Coulomb interactions.\cite{Dietl:2006_a,Ye:2006_a,Kuroda:2007_a} Accordingly, co-doping of DMS with shallow acceptors or donors, during either growth or post-growth processing, modifies $E_{\mathrm{d}}$ and thus provides a mean for the control of ion aggregation. Indeed, the energy of the Coulomb interaction between two elementary charges residing on the nearest neighbor cation sites in the GaAs lattice is 280~meV. This value indicates that the Coulomb interaction can preclude the aggregation, as the gain of energy associated with the bringing two Mn atoms in (Ga,Mn)As is $E_{\mathrm{d}} = 120$~meV.

It is evident that the model in question should apply to a broad class of DMS as well to semiconductors and insulators, in which a constituent, dopant, or defect can exist in different charge states under various growth conditions. As important examples we consider (Ga,Mn)N (Ref.~\onlinecite{Reed:2005_a,Kane:2006_a}) and (Zn,Cr)Te,\cite{Kuroda:2007_a,Ozaki:2005_a,Ozaki:2006_a} in which remarkable changes in ferromagnetic characteristics on co-doping with shallow impurities have recently been reported. In particular, a strong dependence of saturation magnetization $M_{\mathrm{s}}$ at 300~K on co-doping with Si donors and Mg acceptors has been found for (Ga,Mn)N with an average Mn concentration $x_{\mathrm{Mn}} \approx 0.2$\%.\cite{Reed:2005_a} Both double exchange and superexchange are inefficient at this low Mn concentration and for the mid-gap Mn level in question. At the same time, the model of nanocrystal self-organized growth explains readily why $M_{\mathrm{s}}$ goes through a maximum when Mn impurities are in the neutral Mn$^{\mathrm{3+}}$ state, and vanishes if co-doping by the shallow impurities makes all Mn atoms to be electrically charged.

Particularly relevant in this context are data for (Zn,Cr)Te, where strict correlation between co-doping, magnetic properties, and magnetic ion distribution has been put into evidence.\cite{Kuroda:2007_a} It has been found that the apparent Curie temperature $T_{\mathrm{C}}^{\mathrm{(app)}}$ and the aggregation of Cr-rich nanocrystals depend dramatically on the concentration of shallow Nitrogen acceptors in (Zn,Cr)Te.\cite{Ozaki:2005_a,Kuroda:2007_a} Actually, $T_{\mathrm{C}}^{\mathrm{(app)}}$ decreases monotonically when the concentration $x_{\mathrm{N}}$ of nitrogen increases, and vanishes when $x_{\mathrm{Cr}}$ and $x_{\mathrm{N}}$ become comparable. This supports the model as in ZnTe the Cr donor level resides about 1~eV above the nitrogen acceptor state. Accordingly, for $x_{\mathrm{N}} \approx x_{\mathrm{Cr}}$ all Cr atoms become ionized and the Coulomb repulsion precludes the nanocrystal formation, as evidenced by Cr spatial mapping.\cite{Kuroda:2007_a} At the same time, the findings are not consistent with the originally proposed double exchange mechanism,\cite{Ozaki:2005_a} as undoped ZnTe is only weakly $p$-type, so that $T_{\mathrm{C}}$ should be small for either $x_{\mathrm{N}} \approx 0$ and $x_{\mathrm{N}} \approx x_{\mathrm{Cr}}$, and pick at $x_{\mathrm{N}} \approx x_{\mathrm{Cr}}/2$, not at $x_{\mathrm{N}} \approx 0$. Importantly, when native acceptors are compensated by Iodine donor doping, both  $T_{\mathrm{C}}^{\mathrm{(app)}}$ and inhomogeneity in the Cr distribution attains a maximum.\cite{Kuroda:2007_a}

Finally, we mention the case of Mn doped GaAs, InAs, GaSb, and InSb. As argued in the previous section, owing to a relatively shallow character of Mn acceptors and a large Bohr radius, the holes reside in the valence band in these systems. Thus, the Mn atoms are negatively charged, which - according to the model in question -- reduces their clustering, and makes it possible to deposit, by low-temperature epitaxy, a uniform alloy with a composition beyond the solubility limit. Co-doping with shallow donors, by reducing the free-carrier screening, will enhance repulsions among Mn, and allow one to fabricate homogenous layers with even greater $x_{\mathrm{Mn}}$. On the other hand, co-doping by shallow acceptors, along with the donor formation by a self-compensation mechanism,\cite{Yu:2002_a} will enforce the screening and, hence, lead to nanocrystal aggregation.

\section{Outlook}

Over the recent years we have learnt a lot about the effects of solubility limits and self-compensation as well as on the role of the strong coupling occurring in materials with a small bond length, the issues emphasized originally as experimental challenges for the high-temperature hole-controlled ferromagnetism in DMS and DMO.\cite{Dietl:2000_a} A  progress in these fields should lead to  an increase of $T_{\mathrm{C}}$ over 175~K in (Ga,Mn)As,\cite{Wang:2005_a} so-far the highest confirmed value found for uniform tetrahedrally coordinated DMS and DMO.

However, extensive experimental and computational search for multifunctional materials has already resulted in the development of semiconductor and oxide systems  which exhibit surprisingly stable ferromagnetic signatures, often up to well above the room temperature, despite the absence of itinerant holes and a small content of magnetic elements. As discussed here, the ferromagnetism of these compounds, and the associated magnetooptical and magnetotransport functionalities, stem  from the presence of  magnetic nanocrystals embedded coherently in the non-magnetic or paramagnetic matrix. Depending on the host and transition metal combination, the assembled nanocrystals are either metallic or insulating. Importantly, their shape (nanodots {\em vs.} nanocolumns) and size can be altered by growth parameters and co-doping during the epitaxy, allowing one to fabricate {\em in-situ}, {\em e.~g.}, TMR and Coulomb blockade devices.

It is clear that the findings in question open the door for a variety of novel room-temperature applications. In particular, the media in question can be employed for low-power high-density magnetic memories, including spin-injection magnetic random access memories and race-track three dimensional (3D) domain-wall based memories.\cite{Thomas:2007_a}  If sufficiently high TMR will be found, one can envisage the field programmable logic (TMR-based  connecting/disconnecting switches)\cite{Reiss:2006_a} and even all-magnetic logic, characterized by low power consumption and radiation hardness. Furthermore, a combination of a strong magnetic circular dichroism and weak losses suggest possible applications as optical isolators\cite{Amemiya:2006_a} as well as 3D tunable photonic crystals and spatial light modulators for advanced photonic applications.\cite{Park:2002_a} Furthermore coherently grown metallic nanostructures may serve as as a building blocks for all-metallic nanoelectronics and for high quality nanocontacts in nanoelectronics,  optoelectronics, and plasmonics as well as constitute media for thermoelectric applications.\cite{Katayama-Yoshida:2007_b} Worth mentioning is also the importance of hybrid semiconductor/ferromagnetic systems in various proposals of scalable quantum processors.\cite{Loss:1998_a,Dery:2007_a}

\section*{Acknowledgments}
This work was supported in part by NANOSPIN E.~C. project (FP6-2002-IST-015728), by the Humboldt Foundation, and carried out in collaboration with M. Sawicki in Warsaw as well as with the groups of H. Ohno in Sendai, S. Kuroda in Tsukuba, A. Bonanni in Linz, J. Cibert in Grenoble, B. Gallagher in Nottingham,  and L. W. Molenkamp in W\"{u}rzburg.


\end{document}